\input psfig
\documentstyle{l-aa}     

\def\ros{{\sl ROSAT }}
\def\etal{{\it et\,al. }}

\def\ein{{\sl Einstein }}

\def\msun{$M_{\odot}$}

\def\mdot{$\dot M$}
\def\grad{$^\circ$}
\def\it{\sl}
\def\degs{\ifmmode ^{\circ}\else$^{\circ}$\fi}
\def\amin{\ifmmode ^{\prime}\else$^{\prime}$\fi}
\def\asec{\ifmmode ^{\prime\prime}\else$^{\prime\prime}$\fi}
\def\fss{\hbox{$.\!\!^{\rm s}$}}        
\def\farcs{\hbox{$.\!\!^{\prime\prime}$}}  
\def\h{$^{\rm h}$}\def\m{$^{\rm m}$}

\def\rxj{RX~J0527.8--6954}

\begin{document}
 
   \thesaurus{06         
              (11.13.1;  
               02.01.2;  
               13.25.5)  
             }

   \title{The long-term X-ray lightcurve of RX J0527.8--6954}
 
   \author{J. Greiner\inst{1}, R. Schwarz\inst{2}, G. Hasinger\inst{2}, 
        M. Orio\inst{3}\thanks{On leave from Osservatorio Astronomico
        di Torino, 10025 Pino Torinese, Italy}}

   \offprints{J.\,Greiner,\,jcg@mpe-garching.mpg.de}
 
  \institute{$^1$ Max-Planck-Institut f\"ur Extraterrestrische Physik,
             85740 Garching, Germany \\
        $^2$ Astrophysikalisches Institut Potsdam, An der Sternwarte 16, 
             14482 Potsdam, Germany \\
        $^3$ Department of Physics, University of Wisconsin, 1150 Univ. Av.,
             Madison, WI 53706, USA 
       }

   \date{Received October 20, 1995; accepted January 19, 1996}
 
   \maketitle
 
   \begin{abstract}
Supersoft X-ray sources are commonly believed to be stably burning
white dwarfs. However, the observations of some supersoft sources
show dramatic variability of their X-ray flux on timescales ranging from 
days to years.
Here, we present further observational data of the supersoft X-ray source
RX~J0527.8--6954 exhibiting a continuous decline over the past 5 yrs.
With no clear trend of a concordant temperature decrease this might suggest
a evolutionary scenario where the WD leaves the steady burning branch
and the combined effect of reduced luminosity and cooling at constant radius
 produces the observed effect.
 

   \end{abstract}
 
\section{Introduction}

Supersoft X-ray sources (SSS) are characterized by very soft X-ray radiation
of high luminosity. The \ros spectra are well described by
blackbody emission at a temperature of about kT$\approx$ 25--40 eV
and a luminosity close to the Eddington limit (Greiner \etal 1991, Heise 
\etal 1994). After the discovery of supersoft X-ray sources with \ein 
observations, the \ros satellite has discovered more than a dozen new SSS.
Most of these have been observed in nearby galaxies (see Greiner 1995 for
a recent compilation).
Galactic SSS are hard to detect due to the high interstellar absorption
in the galactic plane. Besides the early discovery of the old nova
GQ Muscae as SSS there are now two luminous galactic SSS known with the
classical CAL 83 like properties, namely RX J0925.7--4758 (Motch \etal 1994) and
RX J0019.8+2156 (Beuermann \etal 1995, Greiner \& Wenzel 1995).

The supersoft X-ray source RX~J0527.8--6954 was discovered during the
\ros first light observation (Tr\"{u}mper \etal 1991) of the Large Magellanic 
Cloud (LMC) in June 1990. 
It has an extremely soft X-ray spectrum, with spectral 
parameters (blackbody temperature, absorbing column density) very similar
to CAL 83 (Greiner \etal 1991), the SSS prototype (Long \etal 1981).
Also, it was readily realized that this source must have brightened up
by at least a factor of 10 compared to previous \ein observations
when \rxj\, was in the field of view but not detected.
It was noticed already earlier (Orio and \"Ogelman 1993, Hasinger 1994) 
that the countrate of \rxj\, had decreased substantially since its discovery. 

Here, we use all
available \ros data to document the X-ray variability over the past 5 years.
All the \ros data analysis described in the following has been performed using 
the dedicated EXSAS package (Zimmermann \etal 1994).

\section{Observational results}

\subsection{All-Sky-Survey}

As RX~J0527.8--6954 is close to the south ecliptic pole, it was scanned
during the All-Sky-Survey over a time span of 21 days. The total
observation time resulting from 92 individual scans adds up to 1.96 ksec.

Due to the scanning mode the source has been observed at all possible off-axis
angles with its different widths of the point spread function. For the 
temporal and spectral analysis we have used an 5\amin\, extraction radius 
to ensure that no source photons are missed. No other source down to the 
1$\sigma$ level is within this area. Each photon
event has been corrected for its corresponding effective area. The background
was determined from a circle  13\amin\, off along southern ecliptic
latitude with respect to RX~J0527.8--6954.
The mean count\-rate was determined to (0.14$\pm$0.06) cts/sec.
Due to systematic errors of about 20\% no definite conclusion
can yet be drawn on possible short-term variations of the X-ray flux.

   \begin{table*}
      \caption{Summary of \ros observations covering RX~J0527.8--6954. 
              Given are for each pointing the observation ID (column 1),
              the date of the observation (2), 
              the detector (P=PSPC, H=HRI) (3), the nominal exposure time (4),
              the effective exposure time (5), the total number of counts (6), 
              the off-axis angle of \rxj\, during this observation (7), and
              the distance of \rxj\, to the next rib of the entrance window
              support structure (8).}
            \begin{center}
            \begin{tabular}{lrcrrrrr}
            \hline
            \noalign{\smallskip}
     ROR No. & Date~~~~~ & P/H & T$_{Nom}$ & T$_{Eff}$ & No. of &  Off-axis & 
                                                                 D$_{rib}$~ \\
             &      &     & (sec)     & (sec)     & counts &  angle~  &  \\
            \noalign{\smallskip}
            \hline
            \noalign{\smallskip}
      110173  & June 18, 1990     & P & 2042 & 1418 & 296 & 31\amin & 5\amin \\
      110176  & June 19, 1990     & P & 2168 & 1548 & 292 & 29\amin & 5\amin \\
      110074  & June 20, 1990     & P &  754 &  471 & 72 & 28\amin & 9\amin \\
      110181  & June 21, 1990     & P & 1882 & 1176 & 207 & 46\amin & 15\amin \\
      110090  & June 24, 1990     & P &  457 &  310 & 69 & 25\amin & 3\amin \\
      110234  & July 6, 1990      & H &  779 & 736  & 19 & 13\amin & --~ \\
      110241  & July 7, 1990      & H &  474 &  460 & 12 & 12\amin & --~ \\
      Survey  & Oct. 10--31, 1990 & P & 1965 & 1357 & 189 & 0--55\amin & --~ \\
      141800  & Dec. 11, 1991     & P & 1060 & 541 & 40 & 21\amin & 0\amin \\
      160084  & May 5, 1991       & P & 1757 & 1040 & 100 & 20\amin & 0\amin \\
      300126  & Mar. 5, 1992      & P & 7802 & 4972 & 401 & 22\amin & 0\amin \\
      500004  & Apr. 5, 1992      & P & 1100 & 805 & 47 & 20\amin & 1\amin \\
      400148  & Apr. 6, 1992      & P & 6263 & 6263 & 658 & 1\amin & 20\amin \\
      300172  & May 7--16, 1992   & P & 6371 & 3636 & 403 & 37 & 1\amin \\
      400238  & Nov. 26, 1992     & H & 4063 & 3788 &  38 & 10\amin & --~ \\
      400298  & Dec. 6, 1992      & P & 1058 & 1058 & 61 & 1\amin & 20\amin \\
      300172a & Dec. 16--26, 1992 & P & 2996 & 2163 & 120 & 36\amin & 10\amin \\
      400298a & Mar. 11--16, 1993 & P & 7506 & 7506 & 202 & 1\amin & 20\amin \\
      500141  & Apr. 11, 1993     & P & 5259 & 3709 & 110 & 19\amin & 1\amin \\
      141937  & Apr. 16, 1993     & P & 1979 & 1394 & 55 & 21\amin & 0\amin \\
      141506  & June 16, 1993     & P & 706  & 446 & 15 & 21\amin & 0\amin \\
      300172b & June 14--27, 1993 & P & 3882 &  2630 & 157 & 36\amin & 8\amin \\
      141507  & Aug. 24, 1993     & P & 1334 & 940 & 25 & 22\amin & 0\amin \\
      201689  & Aug. 29/30, 1994  & H & 8463 & 8463 & 16 & 0\amin & --~ \\
      201996  & Aug. 10--12, 1995 & H & 7162 & 7162 &  8 & 0\amin & --~ \\
      600782  & Oct. 22-24, 1995  & H & 12872 & 12588 & 6 & 4\amin & --~ \\
            \noalign{\smallskip}
           \hline
           \end{tabular}
           \end{center}
           \label{poin}
   \end{table*}

For the spectral fitting the X-ray photons in the amplitude channels 11--240 
(though there are almost no photons above channel 50) were binned 
with a constant signal/noise ratio of 5$\sigma$. The fit of a blackbody
model results in an effective temperature of kT$_{\rm bb}$ = 40 eV (with the
absorbing column fixed at its galactic value), very similar to the results 
obtained from fitting the PSPC data of the \ros
first light observation (Greiner \etal 1991).

   \begin{figure*}[htbp]
      \centering{
      \hspace*{.01cm}
      \vbox{\psfig{figure=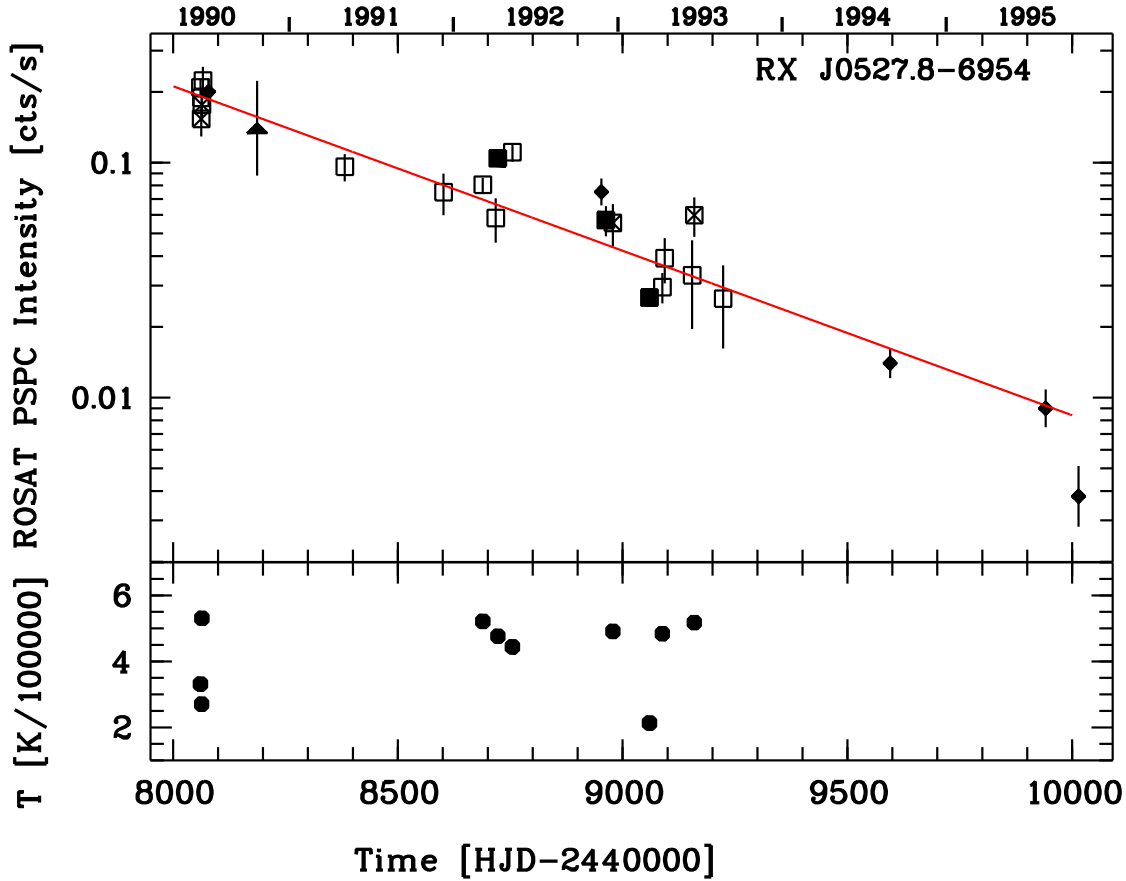,width=15cm,%
          bbllx=1.6cm,bblly=3.2cm,bburx=18.6cm,bbury=16.5cm,clip=}}\par
      }
      \vspace{-0.1cm}
      \caption[longlight]{X-ray lightcurve of \rxj\, over the past 5 years
              as observed with the \ros satellite.
              The triangle marks the mean of the scanning observation during 
              the All-Sky-Survey,
              squares mark PSPC observations (open symbols
              for off-axis angles larger than 15\amin\, and affected by
              the window support structure, crossed squares for
              off-axis angles larger than 15\amin\, and no obscuration
              by ribs, and filled symbols
              for off-axis angles smaller than 15\amin), and filled
              hexagons denote HRI observations  with  the countrates
              transformed to PSPC rates (see text). Systematic errors
              (not included in the plotted error bars) are largest for 
              observations marked with open squares and might reach a factor 
              of 2. The solid line is an exponential with $\tau$=1.7 years.
              The lower panel shows the best-fit blackbody temperature
              (while N$_{\rm H}$ was fixed) derived from the PSPC observations.
              The low temperature point at HJD = 9060 (ROR 400298a) and the 
              spread of best-fit temperatures during the All-Sky-Survey
              are possibly caused by inadequately corrected gain differences
              at various off-axis angles.
              }
         \label{light}
    \end{figure*}

\subsection{PSPC Pointings}

Several pointings with the \ros PSPC have been performed
on \rxj\,, starting with the
first light observation and continuing with several dedicated pointings.
In addition, \rxj\, was also in the field of view of a number of other target 
pointings, mainly within those on the bright supernova remnant N132D which
has been used for calibration purposes. We restricted the analysis to
pointings with effective exposure times larger than 300 sec and
containing \rxj\, at less than 50\amin\, off-axis angle
(Table \ref{poin} gives a complete log of the observations used here).

Depending on the off-axis angle of \rxj\, within the detector, 
photons have been extracted within 6--10\amin\, radius. This extreme size of 
the extraction circle was chosen because
the very soft photons (below channel 20) have a much larger spread in their 
measured detector coordinates.
As usual, the background was determined from a ring well
outside the source without having contaminating sources in there.
Before subtraction, the background photons were normalized to the same 
area as the source extraction circle.
Since RX J0527.8--6954 is affected, unfortunately, by the window
support structure in many pointings, we have developed a dedicated
procedure to correct for the shadowing and wobbling effect. 

As a first step, we took the standard PSPC instrument map together with the 
effective area table to produce an instantaneous correction map. In the second
step these correction maps are added up according to the wobble motion and 
roll-angle using the attitude table. Both these steps were performed
separately in the 11--41 and 42--52 energy channel bands. This energy 
selection is important because the obscuration of sources (scattering)
is energy dependent. We then determine the good time intervals (exposure time)
and after multiplication get two exposure maps (in the separate energy bands)
containing the effects of vignetting and wobbling. As the next step we
determine the relative number of counts of \rxj\, in the 11--41 and 42--52 
bands and use these as weights for summing the two exposure maps.
Finally,
the resulting exposure map is used to compute the effective exposure time 
(given in column 5 of Tab. \ref{poin}) of 
\rxj\, by averaging over the same location and area as source photons have been
extracted. 

\subsection{HRI Pointings}

There are also a number of HRI pointings which cover \rxj, namely two 
pointings during the verification phase, one pointing performed in November 
1992 in the framework of LMC X-ray source identifications (see 
Cowley \etal 1993) and three dedicated, on-axis pointings in August 1994,
August 1995 and October 1995
(see Tab. \ref{poin}). We restricted the analysis to pointings with 
effective exposure times longer than 300 sec and at less than 15\amin\,
off-axis angle (excluding two verification phase pointings).
Source photons have been extracted within 2--3\amin\, and were
background and vignetting corrected.

In order to compare the HRI intensities of \rxj\, with those measured with the 
PSPC we determined the PSPC/HRI countrate ratio for this supersoft X-ray 
spectrum in the following two ways:
1) In an empirical approach we selected a few single white dwarfs (WDs) which 
have been observed with both, the PSPC and HRI, and derive a conversion factor
PSPC/HRI = 7.8. This is thought to be an upper limit because such isolated
WDs are less absorbed than SSS and thus sample even the lowest PSPC 
channels.
2) We have used the best fit model of the PSPC spectrum 
as input for computing the expected HRI countrate using its up-to-date 
response matrix and effective area. The result is sensitive to the
temperature chosen and gives a ratio of PSPC/HRI = 7.7 (7.0) for a
blackbody temperature of kT = 35 (40) eV. 
Thus, we adopt a ratio of 7.5, and the converted countrates of the HRI
pointings are also included in Fig. \ref{poin}.

We derive a best-fit position ($\pm$5\asec)
of R.A. (2000.0) = 05\h27\m48\fss9, Decl. (2000.0) = --69\grad54\amin09\asec\,
which results from the position averaging of the two individual on-axis HRI 
pointings in August 1994 and August 1995 (which differ by 4\asec). 
The averaging reduces the 
irreproducible r.m.s. scatter of $\approx$5\asec\, from the individual 
pointings due to the fact that the roll angles are different. 
This new position differs by only 2\farcs5 from that given in Hasinger 
(1994) which arose from the averaging of several off-axis (4 PSPC and 1 HRI)
pointings. Our new position strengthens the conjecture of Hasinger (1994)
that the blue object proposed by Cowley \etal (1993) as a strong
counterpart candidate is too distant to be a likely counterpart.
Though this new position is only 6\asec\, from the Cowley
\etal (1993) position, it is in the opposite direction with respect to the
above mentioned blue object.

\subsection{The lightcurve}

Fig. \ref{light} shows the X-ray lightcurve of \rxj\, (i.e. the background
subtracted counts divided by the effective exposure time) as deduced from 
19 \ros PSPC pointings, six HRI pointings and the All-Sky-Survey data between
1990 and 1994. Two main features can be recognized immediately from this 
overall 5 year lightcurve: 1) The source has exponentially declined in X-ray 
intensity since its first \ros observations in 1990 ($\tau$=1.7$\pm$0.1 yrs), 
and 2) there is considerable scatter in the decline which is larger than our 
estimate of the remaining systematic errors in correcting for the effects
of the window support structure.

Surprisingly, the last HRI observation in October 1995 shows \rxj\, to be 
considerably fainter than the extrapolation of the exponential decline. 
The X-ray intensity has dropped by a factor of 2.5 within two months.
We have checked the housekeeping data of this pointing for any anomalous
behaviour in detector properties or background variation - with negative
result. Unfortunately, the only steady X-ray source (the SNR N132D) is
at 20\amin\, off-axis and thus cannot be used as an observational calibration.
With the data products of this observation being so recent, further
checks of the instrument performance are certainly necessary.
If this intensity drop indeed is real, one might speculate on having
caught the source during the switch-off.

The total X-ray amplitude between maximum and minimum observed \ros 
intensity is a factor of 50 within these 5 years. This is about a factor 
five larger than the estimated amplitude deduced from the \ein non-detection.
At the present X-ray intensity (and during all the last two years)
the source would have been invisible again for the \ein observatory.
Thus, the variability timescale of \rxj\, is less than $\approx$10--15 years. 
Due to the lack of observations inbetween, it is also conceivable that we 
caught the source in June 1990 just during its onset. Therefore, the
variability timescale could be as short as five years.

Though the number of counts detected during the individual PSPC pointings is
mostly rather low, we investigated the possibility of X-ray spectral changes
during the decline. First, we kept the absorbing column fixed at its galactic
value and determined the temperature being the only fit parameter. We find
no systematic trend of a temperature decrease (lower panel of Fig. \ref{light}).
Second, we kept the temperature fixed (at 40 eV in the first run and at the
best fit value of the two parameter fit in the second run) and checked 
for changes in N$_H$, again finding no correlation.

\section{Discussion}

The most popular model of SSS involves steady nuclear burning on the surface of
an accreting WD (van den Heuvel \etal 1992).
If \.{M} exceeds a certain value that depends on the WD
 mass and other physical parameters, hydrogen burning
 on the WD surface is stable.

There are at least three possible phenomena which might explain completely or
partially the X-ray variability of \rxj:
\begin{enumerate}
\vspace*{-0.15cm}
\item Changes in \mdot within the small range which is necessary for stable 
hydrogen burning. The luminosity  amplitude in this case can be only a factor
2.3--2.7 (see Fig. 9 of Fujimoto 1982 and Figs. 2 and 8 of Iben 1982). 
However, since the nuclear 
luminosity depends mainly on the conditions of the burning envelope, 
the corresponding changes are expected on the accretion time scale, i.e. 
much longer than observed in \rxj.
\item The atmospheric layers are expanding and thus the effective
temperature cools, while the bolometric luminosity remains
constant (evolution along the horizontal track in the log(T)-
log(L) plane). The expansion could be caused by increased mass
transfer from the secondary. This mechanism has already been suggested
by Pakull \etal (1993) for RX J0513.9--6951. 
\item The WD is cooling at constant radius. Due to the strong sensitivity 
of the \ros countrate on the temperature the X-ray amplitude of a factor of 
10 does not necessarily translate into a factor of 10 change in bolometric 
luminosity (even if the latter is assumed to be dominated by the soft 
X-rays). Thus, the WD can remain within the stable burning range
(with its factor 2--3 reduction in bolometric luminosity) while
the concordant temperature decrease will shift the Wien tail out of the
\ros window.
If the cooling alone would account for the X-ray intensity decay, then
the observed intensity amplitude corresponds to a temperature amplitude
of at least two (depending on the absolute temperature). This translates into
a cooling time of the order of five years. We can assume that
the shortest time for the white dwarf cooling 
is only slightly longer than the time necessary to burn all the hydrogen
envelope mass   once there is no mass transfer at all,
like it is usually supposed after a nova outburst (Starrfield, private
communication). This can be as short as 5 years for a white
dwarf of 1 $M_\odot$ (Kato \& Hachisu, 1994).
\vspace*{-0.1cm}
\end{enumerate}

Possibility 1 seems to be ruled out as the only cause of the exponential decline
due to the observed intensity amplitude as well as the timescale.
Though we cannot clearly distinguish yet between possibilities 2 
and 3, the lack of a clear trend of a temperature decrease would favour 
possibility 3. In addition, scenario 2 only works if the accretion rate is
near the Eddington limit for a burning WD.

Alternatively, the gradual decline of the X-ray intensity also might suggest
that we observe the decay phase after a shell flash. Thus, it is important
to ask when the  decay of \rxj\, might have started. The recurrence time of 
hydrogen flashes  is inversely proportional to the mass accretion
rate and to the white dwarf mass. 
Since \rxj\, has not been detected with \ein observations,
the shell flash must have occurred after 1981. Moreover, since with \ros
we witness a decline from the beginning,
 the hydrogen-burning plateau phase (horizontal track in the H--R
diagram) must have been short.
Thus, even if the
shell flash happened just after the \ein observation then it would still
need a considerable fine tuning of WD mass and accretion rate 
($\dot M \leq  10^{-8} M_\odot$ yr$^{-1}$) for the WD to be
on the declining portion of the evolutionary track within only 10 years.
Moreover, a shell flash with such properties would be accompanied by an
optical brightening of the object to M$_{\rm V}\simeq$--5. At the LMC
distance this would correspond to a visual brightness of 14th magnitude
which very certainly would not have passed undetected.

It seems more likely that the mechanism causing the variability of \rxj\,
was a hydrogen flash on a massive white dwarf (m$\geq
1.1 M_\odot$) accreting at a high rate
(10$^{-7}$ \msun/yr) as described in Iben (1982) and Fujimoto (1982).
In an empirical model of recurrent SSS, Kahabka (1995) relates the envelope 
mass to the decay and recurrence times of these sources. The latter can 
be used in turn as observables
to determine the WD mass and its accretion rate. 
From a preliminary analysis of the X-ray decline of \rxj\, Kahabka (1995)
adopted a value of 5 yrs for the time for return to minimum after a flash,
and a recurrence time of 10 yrs. These numbers have not changed with
the herewithin reported results. 
We note in passing, that the time for return to minimum does not correspond to
$\tau$ in the exponential exp(-t/$\tau$) though the 
recurrence time can also be of the order of 5 yrs. With these two
timescales Kahabka (1995) derived 
M$_{\rm WD}$=1.14--1.34 \msun\, and \mdot$_{\rm accr}$=1.2--3.4$\times$10$^{-7}$
\msun/yr. The rather high WD mass is also compatible with the temperature
of 5--6$\times$10$^5$ K and the derived luminosity.

\section{Summary}

\rxj\, has been observed to exponentially decline in X-ray intensity since 
its discovery in June 1990 suggesting an X-ray timescale
of the order of 10--15 years. Given the strong sensitivity of the \ros
detectors on small variations in the temperature of the object, 
the amplitude of decline (factor 50 within several years)
is compatible with temperature changes of the WD.
The rather rapid timescale of the decline suggests a massive WD.

The rapid drop in X-ray intensity of \rxj\, within two months at the end
of 1995 adds even more importance in continuing the X-ray monitoring of
this enigmatic source.

\begin{acknowledgements}
JG is supported by the Deutsche Agentur f\"ur
Raumfahrtangelegenheiten (DARA) GmbH under contract FKZ 50 OR 9201.
We are extremely grateful to Y.-H. Chu for providing the data of the
October 1995 HRI observation.
We thank the referee C. Motch for valuable comments.
The \ros project
is supported by the German Bundes\-mini\-ste\-rium f\"ur Bildung,
Wissenschaft und Forschung (BMBF/DARA) and the Max-Planck-Society.
\end{acknowledgements}

\end{document}